\RequirePackage{lineno}

\documentclass[aps,prc,showpacs,preprintnumbers,amsmath,amssymb,superscriptaddress]{revtex4}

\usepackage{graphicx} 
\usepackage{amssymb}
\usepackage{dcolumn}
\usepackage{bm}
\usepackage{color}

\newcommand{ \be }{\begin{linenomath*}\begin{eqnarray}}
\newcommand{ \ee }{\end{eqnarray}\end{linenomath*}}
\newcommand{ \la }{\langle}
\newcommand{ \lla }{\left\langle}
\newcommand{ \ra }{\rangle}
\newcommand{ \rra }{\right\rangle}
\def \mean#1 {{\la #1 \ra}}

\definecolor{dgreen}{cmyk}{1.,0.,1.,0.2}        
\definecolor{orange}{cmyk}{0.,0.353,1.,0.}    

\begin{document}
\title{Identity Method Revisited}
\author{Claude A. Pruneau}
\affiliation{Department of Physics and Astronomy, Wayne State University}
\date{\today}

\begin{abstract}
We discuss the impact of finite particle losses associated with instrumental effects in  measurements of moments of produced multiplicities with the Identity Method towards the evaluation of fluctuation measures such as $\nu_{dyn}$. We show that the identity method  remains applicable provided it is  modified to  determine factorial moments $\la N(N-1)\ra$, rather than moments $\la N^2\ra$. We further show that  $\nu_{dyn}$  remains robust if detection
efficiencies are uniform across the measurement's acceptance. The robustness is lost, however, if detection efficiencies are momentum dependent, although the identity methods
remains applicable  provided detection efficiencies  can be determined with sufficient accuracy. 
\end{abstract}
\maketitle

\section{Introduction}

Studies of fluctuations of the relative yield of produced particles  in high-energy nucleus-nucleus collisions provide valuable information on the particle production dynamics, the collision system evolution,  and might also enable the identification of anomalous  behavior signaling deconfinement or the existence of critical behavior~\cite{PhysRevLett.81.4816,PhysRevLett.95.182301,PhysRevC.72.064903}. Measurements of integral correlations based on the $\nu_{dyn}$ fluctuation measure~\cite{Voloshin:1999yf,Pruneau:2002yf}, in particular, have received a growing level of interest because this observable provides several advantages experimentally and phenomenologically. It is indeed straightforward  to measure thanks to its  rather simple definition based on a combination of ratios of second factorial moments to the square of inclusive averages, and because it is nominally robust against particle losses due to instrumental effects. It is also relatively insensitive to collision volume uncertainties and fluctuations and its phenomenological interpretation is thus relatively straightforward. 

The $\nu_{dyn}$ fluctuation measure has been used to study net-charge fluctuations~\cite{PhysRevC.68.044905,PhysRevC.79.024906,Abelev:2012pv} as well as fluctuations of the relative yield of different particle species~\cite{Abelev:2009ai,Kumar:2013cqa}. Measurements of relative species yield fluctuations typically utilize conventional particle selection techniques based on measurements of specific energy loss and time-of-flight measurements.  In the context of this technique, particles must be identified and counted event-by-event to determine the number (multiplicity) of particles  of each species of interest, and calculate their first  and  second factorial moments within 
the collision dataset. Evidently, measurements of specific energy loss or time-of-flight provide unambiguous particle identification capabilities  only across a rather limited kinematic range. Beyond such a range,  considerable PID ambiguity typically arises. Ambiguity and signal contamination may be suppressed by using narrower selection cuts but these usually imply significant reductions in detection efficiency. 
In an effort to avoid signal contamination, ambiguities, and efficiency losses implied by narrow PID selection criteria,  authors of Refs.~\cite{Gazdzicki:1997gm,Rustamov:2012bx,Gorenstein:2011hr}  have developed a technique known as {\it identity method} which relies on the probability a given particle might be of a given type or species based on the value of the PID signal and the estimated line shape of such signal for distinct particle species. The method is straightforward for measurements of single particle spectra but becomes significantly more complicated for the evaluation of second or higher moments of multiplcities. Be it as it may, Ref.~\cite{Gorenstein:2011hr} presents a well defined and relatively straightforward method for the evaluation of second moments and covariances. The method is quite elegant but unfortunately neglects effects associated with particle losses. It is the purpose of this work to investigate the impact of such losses  and whether the method can be modified to account for them. 

This paper is divided as follows. Section~\ref{sec:EfficiencyLoss} presents a brief review of the impact of uncorrelated efficiency losses in cases where particle counting is unambiguous and exact. Section~\ref{sec:IdentityMethod} builds on the identity method described in Refs.~\cite{Gazdzicki:1997gm,Gorenstein:2011hr} and presents a discussion of the impact of uncorrelated particle losses on the calculation of the moments of the identity variables $W_p$. 
The method is further expanded in sec.~\ref{sec:IMwithPtBins} to account for momentum dependent efficiencies. This work is summarized in sec.~\ref{sec:Summary}. 

\section{Measuring Multiplicity Moments in the presence of efficiency losses}
\label{sec:EfficiencyLoss}
We formulate the discussion in the context of a measurement of the $\nu_{dyn}$ observable but the results presented in this work can be straightforwardly extended to other fluctuation observables. The observable  $\nu_{dyn}$ and its properties were introduced and discussed in detail in Ref.~\cite{Pruneau:2002yf}:
\be\label{eq:nudyn}
\nu_{dyn}(N_p,N_q) = \frac{\la N_p\left( N_p-1\right)\ra }{\la N_p\ra^2 } +  \frac{\la N_q\left( N_q-1\right)\ra }{\la N_q\ra^2 } - 2  \frac{\la N_p N_q\ra }{\la N_p\ra \la N_q\ra },
\ee
The variables $N_p$ and $N_q$ represent the multiplicities of produced particles, of species of type $p$ and $q$, respectively, measured event-by-event, within the fiducial volume $\Omega$ of the experiment.  More generally, one is interested in measuring factorial and cross moments of  multiplicities $N_p$ and $N_q$  of particle  species $p$ and $q$, with $p,q=1,\ldots, K$ denoting $K$ distinct particle species (e.g., $1=$pion, $2=$kaon, $3=$proton), observable and countable event-by-event. These moments  are determined by the joint probability of the $K$ particle species, which we denote $p_{\rm T}(N_1,N_2,\ldots,N_K)$:
\be\nonumber
\la N_p\ra &\equiv & \sum_{N_p=0}^{\infty}  P_{\rm T}(N_1,\ldots,N_p,\ldots, N_K) N_p, \\ \label{eq:ProdMultMoments}
\la N_p^2\ra &\equiv & \sum_{N_p=0}^{\infty}  P_{\rm T}(N_1,\ldots,N_p,\ldots, N_K) N_p^2, \\ \nonumber
\la N_p N_q\ra &\equiv & \sum_{N_p=0}^{\infty}  P_{\rm T}(N_1,\ldots,N_p,\ldots, N_q, \ldots, N_K) N_p N_q. 
\ee
Evidently, not all produced particles are properly counted given there are instrumental losses. We label the  multiplicity  of measured (detected) particles using lower case letters, $n_p$. The instrumental losses are modeled with independent  binomial distributions, $B(n_p|N_p,\varepsilon_p)$, $p=1,\ldots, K$, which we write
\be \label{eq:BinomialDist}
B(n_p|N_p,\varepsilon_p) = \frac{N_p!}{n_p! (N_p-n_p)!} \varepsilon_p^n \left( 1 - \varepsilon_p\right)^{N_p-n_p},
\ee
where $\varepsilon_p$ represent the detection efficiency of particle species $p$. In general, the efficiencies $\varepsilon_p$ differ for species $p=1,\ldots, k$. 
The joint probability $P_{\rm M}(n_1,n_2,\ldots,n_K)$ of the number of observed particles is obtained by summing over all multiplicities the product of the joint probability of produced multiplicities $p_{\rm T}(N_1,N_2,\ldots,N_K)$ by the probabilities of observing the multiplicities $n_p$ given the produced multiplicities $N_p$.
\be\label{eq:Pm}
P_{\rm M}(n_1,n_2,\ldots,n_K) = \sum_{N_1=0}^{\infty} \sum_{N_2=0}^{\infty} \cdots \sum_{N_K=0}^{\infty} P_{\rm T}(N_1,N_2,\ldots,N_K)B(n_1|N_1,\varepsilon_1)B(n_2|N_2,\varepsilon_2)\times \cdots \times B(n_K|N_K,\varepsilon_K).
\ee
The moments of the observed multiplicities are then calculated similarly as those of the produced multiplicities and one gets
\be\nonumber
\la n_p\ra &\equiv & \sum_{n_p=0}^{\infty}  P_{\rm M}(n_1,\ldots,n_p,\ldots,n_K) n_p, \\ \label{eq:MeasMultMoments}
\la n_p^2\ra &\equiv & \sum_{n_p=0}^{\infty}  P_{\rm M}(n_1,\ldots,n_p,\ldots,n_K) n_p^2, \\ \nonumber
\la n_p n_q\ra &\equiv & \sum_{n_p=0}^{\infty}  P_{\rm M}(n_1\ldots,n_p,\ldots, n_q, \ldots,n_K) n_p n_q. 
\ee
It is then straightforward to verify (see for instance Ref.~\cite{Pruneau:2002yf}) that the moments of the observed multiplicities are related to those of the produced multiplicities according to 
\be\nonumber
\la n_p \ra &=& \varepsilon_p \la N_p \ra \\  \label{eq:MeasVsProdMoments}
\la n_p^2 \ra &=& \varepsilon_p \left( 1-\varepsilon_p\right) \la N_p \ra + \varepsilon_p^2 \la N_p^2 \ra \\  \nonumber
\la n_p n_q\ra &=& \varepsilon_p \varepsilon_q \la N_p N_q \ra.
\ee
and the measured factorial moments $\la n_p\left( n_p - 1\right)\ra $ are 
\be\label{eq:FacMeasVsProdMoments}
\la n_p\left( n_p - 1\right)\ra =  \varepsilon_p^2 \la N_p\left( N_p - 1\right)\ra.
\ee
The observable $\nu_{dyn}$ is thus considered robust because efficiencies for species $p$ and $q$ cancel out of each of the three terms of Eq.~\ref{eq:nudyn}.

The neglect of particle losses can have a significant impact on measurements of fluctuations. To illustrate this impact, we assume that the average multiplicity of species $p$ is of order $\la N_p\ra = 100$ with a variance of $\la \Delta N_p^2\ra = 90$. The second moment of $N_p$ is thus $\la N_p^2\ra = 10,090$, and $\la \Delta N_p^2\ra/\la N_p\ra^2=0.009$. Assuming the efficiency is $\varepsilon_p = 0.8$, we find, using Eq.~(\ref{eq:MeasVsProdMoments}), $\la n_p\ra = 80$, $\la n_p^2\ra = 6,473.6$, and $\la \Delta n_p^2\ra/\la n_p\ra^2=0.0115$, which amounts to a 28\% error. However, one verifies that $\la n(n-1)\ra/\la n_p\ra^2 = \la N(N-1)\ra/\la N_p\ra^2$ holds perfectly. We thus expect that to the extent that the identity method enables proper unfolding of the PID signal line shape, the moments $\la n\ra$ and  $\la n^2\ra$ shall then be heavily biased by particle losses, but quantities such as $\la n(n-1)\ra/\la n_p\ra^2$ shall remain robust and unbiased, that is, independent of particle detection efficiencies. We show in the next section that this conclusion holds if  the efficiencies are momentum independent.

\section{The Identity Method}
\label{sec:IdentityMethod}

The identity method was introduced in Ref.~\cite{Gazdzicki:1997gm} for two species, $p=1,2$, and extended in Ref.~\cite{Rustamov:2012bx,Gorenstein:2011hr} for $K>2$ species, i.e., for $p,q=1,\ldots, K>2$, and the determination of higher moments. 
It is based on the realization that it is often not possible, experimentally, to uniquely identify a particle species based on observables such as average energy loss 
in the gas of a Time Projection Chamber, time-of-flight measurement, or any other techniques aiming at the determination of the mass of measured particles. Indeed, one finds, in general, that there are limited kinematic regimes in which different species can be unambiguously identified (i.e., identified with perfect certainty)  based on a particle identification (PID) observable, $s$. In most situations and kinematic ranges,  however, there remains varying degrees of ambiguity. For instance, a given particle might likely be a pion, but there might be a finite probability that it is a kaon or a proton instead. This leads to contamination of the moments $\la N_q\ra$ and $\la N_q(N_q-1)\ra$ which may have a rather detrimental impact on the evaluations of correlation observables such as $\nu_{dyn}$. Within the identity method, rather than summing integer counts (e.g., increasing a counter by one unit if the particle is a pion, and zero, otherwise) and neglecting such contaminations, one accounts for ambiguities by summing  weights $\omega_p(s)$ for each PID hypothesis. The weights are determined particle-by-particle, and for each hypothesis $p$,  according to the probability density of observing $s$ in the range $[s,s+ds]$, for species $p$
\be\label{eq:pidProb}
\omega_p(s) \equiv \frac{\rho_p(s)}{\rho(s)},
\ee
where $\rho(s)\equiv\sum_{i=1}^K \omega_i(s)$. The functions $\rho_p(s)$ represent the {\bf line shapes} of the PID signal $s$ for species $p=1,\ldots, K$, determined from an average over a large ensemble of events. Several types of  PID signal $s$ may be used, including the average energy loss of a particle determined in an ionization chamber (e.g., a Time Projection Chamber), a particle's time of flight or mass determined from a combination of other observables, etc. One   defines an event-by-event variable $W_p$, hereafter called identity variable for species $p$, as the sum of the  weights $\omega_p(s_i)$ calculated for all $M$ particles of an event:
\be\label{eq:IdentityVariable}
W_p = \sum_{i=1}^{M} \omega_p(s_i).
\ee
The identity method   involves the calculation of the moments of the identity variables $\la W_p\ra$, $\la W_p^2\ra$, and $\la W_p W_q\ra$, for all relevant species $p$ and $q$, and  from which the moments  $\la N_p\ra$, $\la N_p^2\ra$, and $\la N_p N_q\ra$ can be nominally extracted by solving a linear equations derived in Ref. \cite{Gorenstein:2011hr}. However, the identity method as outlined in Ref. \cite{Gorenstein:2011hr} neglects the detector response and  does not account for particle losses. Extracted multiplicity moments $\la N_p\ra$ and $\la N_p^2\ra$  are consequently biased and the results obtained may thus be unreliable. This oversight is easily remedied and we derive, in this and following section, formula for the extraction of moments that include effects associated with  efficiency losses. 
  
Toward this end, we proceed to calculate the expectation value of the moments $\la W_p\ra$, $\la W_p^2\ra$, and $\la W_p W_q\ra$ and show they can be related to the expectation value of the moments $\la N_p\ra$, $\la N_p^2\ra$, and $\la N_p N_q\ra$ even in the presence of particles losses. We shall however need efficiencies $\varepsilon_p$,  defined by Eq.~(\ref{eq:BinomialDist}), for each of the particle species $p$ of interest. In general, in a given event, there shall be $n_1$ particles of type $1$, $n_2$ particles of type $2$, and so forth. Assuming there are $K$ species of interest, the variable $W_p$  may then be written
\be\nonumber
W_p &=& \sum_{i_1=1}^{n_{1}} \omega_p(s^{(1)}_{i_1}) + \sum_{i_2=1}^{n_{2}} \omega_p(s^{(2)}_{i_2})+ \cdots +  \sum_{i_K=1}^{n_{K}} \omega_p(s^{(K)}_{i_K}), \\  \label{eq:IdentityVariableCalc}
&=&  \sum_{j=1}^{K}\sum_{i_j=1}^{n_{j}} \omega_p(s^{(j)}_{i_j}),
\ee
which includes $K$ distinct sums consisting of $n_1$, $n_2$, $\ldots$, and $n_K$ terms. The variables $s^{(j)}_{i_j}$ represent the PID variables that might be observed for particles of species  $j$. In order to calculate the moments, one must sum over all permissible permutations of the multiplicities $n_1$, $n_2$, $\ldots$, and $n_K$ and all possible values of the variables  $s^{(j)}_{i_j}$. Expressing the joint probability $P_{\rm M}(n_1,n_2,\ldots,n_K)$ 
according to Eq.~(\ref{eq:Pm}), the expectation value of $\la W_p\ra$ may then be written
\be\label{eq:IdentityVariableFirstMoment}
\la W_p\ra &=& \sum_{n_1=0}^{N_1}\sum_{n_2=0}^{N_2}\cdots \sum_{n_K=0}^{N_K}
\sum_{N_1=0}^{\infty}\sum_{N_2=0}^{\infty}\cdots \sum_{N_K=0}^{\infty} 
P_{\rm T}(N_1,N_2,\ldots,N_K) \\ \nonumber
 & & \times B(n_1|N_1,\varepsilon_1)B(n_2|N_2,\varepsilon_2)\times \cdots \times B(n_K|N_K,\varepsilon_K) \\ \nonumber
 & & \times\prod_{i_1=1}^{n_1} \int p_1(s^{(1)}_{i_1}) ds^{(1)}_{i_1} \times \prod_{i_2=1}^{n_2} \int p_2(s^{(2)}_{i_2}) ds^{(2)}_{i_2}
 \times \cdots \times \prod_{i_K=1}^{n_K} \int p_K(s^{(K)}_{i_K}) ds^{(K)}_{i_K}\\ \nonumber
 & & \times \left[ \sum_{j=1}^{K}\sum_{i'_j=1}^{n_{j}} \omega_p(s^{(j)}_{i'_j}) \right],
\ee
where the functions $p_p(s^{(p)}_{i_p}) \equiv \rho_p(s^{(p)}_{i_p})/\la n_p\ra$ represent the probability density of 
observing PID variable values $s^{(p)}_{i_p}$. Evaluation of the above expression is accomplished by distributing the $K$ terms of $W_p$ and changing the order of the sums:
\be \nonumber
\la W_p\ra &=& \sum_{i'_1=1}^{n_{1}}  \sum_{N_1=0}^{\infty}\sum_{N_2=0}^{\infty}\cdots \sum_{N_K=0}^{\infty} \sum_{n_1=0}^{N_1}\sum_{n_2=0}^{N_2}\cdots \sum_{n_K=0}^{N_K} 
P_{\rm T}(N_1,N_2,\ldots,N_K) \\ \nonumber
 & & \times B(n_1|N_1,\varepsilon_1)B(n_2|N_2,\varepsilon_2)\times \cdots \times B(n_K|N_K,\varepsilon_K) \\ \nonumber
 & & \times \prod_{i_1=1}^{n_1} \int  p_1(s^{(1)}_{i_1}) ds^{(1)}_{i_1} \times \prod_{i_2=1}^{n_2} \int p_2(s^{(2)}_{i_2}) ds^{(2)}_{i_2} 
 \times \cdots \times \prod_{i_K=1}^{n_K} \int p_K(s^{(K)}_{i_K}) ds^{(K)}_{i_K} \times \omega_p(s^{(1)}_{i'_1}) \\ \nonumber
 & &+  \sum_{i'_2=1}^{n_{2}}  \sum_{N_1=0}^{\infty}\sum_{N_2=0}^{\infty}\cdots \sum_{N_K=0}^{\infty} \sum_{n_1=0}^{N_1}\sum_{n_2=0}^{N_2}\cdots \sum_{n_K=0}^{N_K} 
p_{\rm T}(N_1,N_2,\ldots,N_K) \\ \nonumber
 & & \times  B(n_1|N_1,\varepsilon_1)B(n_2|N_2,\varepsilon_2)\times \cdots \times B(n_K|N_K,\varepsilon_K) \\ \nonumber
 & & \times \prod_{i_1=1}^{n_1} \int  p_1(s^{(1)}_{i_1}) ds^{(1)}_{i_1} \times \prod_{i_2=1}^{n_2} \int p_2(s^{(2)}_{i_2}) ds^{(2)}_{i_2}
 \times \cdots \times \prod_{i_K=1}^{n_K} \int p_K(s^{(K)}_{i_K}) ds^{(K)}_{i_K} \times \omega_p(s^2_{i'_2})\\ \nonumber
 & & \cdots \\ \nonumber
 & & +  \sum_{i'_K=1}^{n_{k}}  \sum_{N_1=0}^{\infty}\sum_{N_2=0}^{\infty}\cdots \sum_{N_K=0}^{\infty} \sum_{n_1=0}^{N_1}\sum_{n_2=0}^{N_2}\cdots \sum_{n_K=0}^{N_K} 
p_{\rm T}(N_1,N_2,\ldots,N_K) \\ \nonumber
 & & \times   B(n_1|N_1,\varepsilon_1)B(n_2|N_2,\varepsilon_2)\times \cdots \times B(n_K|N_K,\varepsilon_K) \\ \nonumber
 & & \times \prod_{i_1=1}^{n_1} \int  p_1(s^{(1)}_{i_1}) ds^{(1)}_{i_1} \times \prod_{i_2=1}^{n_2} \int p_2(s^{(2)}_{i_2}) ds^{(2)}_{i_2}
 \times \cdots \times \prod_{i_K=1}^{n_K} \int p_K(s^{(K)}_{i_K}) ds^{(K)}_{i_K} \times \omega_p(s^{(K)}_{i'_K}).
\ee 
Integrals of the form $\int  p_p(s) ds$  yield  unity by definition of the probabilities $p_p(s)$. Introducing, 
\be
u_{pq} =  \int p_q(s)  \omega_p(s) ds,
\ee
and carrying first the sums on observed multiplicities and next those on produced multiplicities, one gets
\be \nonumber
\la W_p\ra &=&  \sum_{N_1=0}^{\infty}\sum_{N_2=0}^{\infty}\cdots \sum_{N_K=0}^{\infty} 
P_{\rm T}(N_1,N_2,\ldots,N_K) \left[ u_{p1} \varepsilon_1 N_1 +  u_{p2}  \varepsilon_2  N_2 + \cdots +  u_{pK} \varepsilon_K  N_K \right], \\ \label{eq:WpFirstMom}
& = & \sum_{i=1}^K  u_{pi} \varepsilon_i \la N_i\ra
\ee
The coefficients $u_{pi} \varepsilon_i$ nominally form a $K \times K$ square matrix  that  can be  
inverted to solve for the moments $\la N_i\ra$. However, this requires a priori knowledge of the efficiencies $\varepsilon_i$. It is thus more convenient to factor  the efficiencies out of the matrix and define uncorrected  multiplicities $N'_p=\varepsilon_p N_p$. The corrected moments $\la N_i \ra$ can  then be estimated with
\be\label{eq:Nmom}
\la N_i \ra = \frac{\la N'_i \ra}{\varepsilon_i} = \frac{ \left({\bf U}^{-1} \la \vec W\ra\right)_i}{\varepsilon_i}
\ee
where ${\bf U}$ is a square matrix, with elements $u_{pi}$.

Calculation of the  second moments $\la W_p^2\ra$ proceeds similarly but one must expand 
the square of $W_p$ in terms of sums over single particle particle species and pairs of species: 
\be\nonumber
\la W_p^2\ra &=& \sum_{n_1=0}^{N_1}\sum_{n_2=0}^{N_2}\cdots \sum_{n_K=0}^{N_K}
\sum_{N_1=0}^{\infty}\sum_{N_2=0}^{\infty}\cdots \sum_{N_K=0}^{\infty} 
P_{\rm T}(N_1,N_2,\ldots,N_K) \\ \nonumber
 & & \times B(n_1|N_1,\varepsilon_1)B(n_2|N_2,\varepsilon_2)\times \cdots \times B(n_K|N_K,\varepsilon_K) \\  \nonumber
 & & \times\prod_{i_1=1}^{n_1} \int p_1(s^{(1)}_{i_1}) ds^{(1)}_{i_1} \times \prod_{i_2=1}^{n_2} \int p_2(s^{(2)}_{i_2}) ds^{(2)}_{i_2}
 \times \cdots \times \prod_{i_K=1}^{n_K} \int p_K(s^{(K)}_{i_K}) ds^{(K)}_{i_K}\\ \nonumber
 & & \times \left[ \sum_{j=1}^{K}\sum_{i'_j=1}^{n_{j}} \omega_p(s^{(j)}_{i'_j}) \right]^2, \\ \label{eq:Wp2Detail} 
&=& \sum_{n_1=0}^{N_1}\sum_{n_2=0}^{N_2}\cdots \sum_{n_K=0}^{N_K}
\sum_{N_1=0}^{\infty}\sum_{N_2=0}^{\infty}\cdots \sum_{N_K=0}^{\infty} 
p_{\rm T}(N_1,N_2,\ldots,N_K) \\ \nonumber
 & & \times B(n_1|N_1,\varepsilon_1)B(n_2|N_2,\varepsilon_2)\times \cdots \times B(n_K|N_K,\varepsilon_K) \\ \nonumber
 & & \times\prod_{i_1=1}^{n_1} \int p_1(s^{(1)}_{i_1}) ds^1_{i_1} \times \prod_{i_2=1}^{n_2} \int p_2(s^{(2)}_{i_2}) ds^{(2)}_{i_2}
 \times \cdots \times \prod_{i_K=1}^{n_K} \int p_K(s^{(K)}_{i_K}) ds^{(K)}_{i_K}\\ \nonumber
 & & \times \left\{ 
    \sum_{j=1}^K \sum_{i'_j=1}^{n_{j}} \left[\omega_p(s^{(j)}_{i'_j})\right]^2 
    + \sum_{j=1}^K \sum_{i'_1\ne i{''}_1=1}^{n_{j}} \omega_p(s^{(j)}_{i'_j})  \omega_p(s^{(j)}_{i{''}_{j}}) 
     +  \sum_{j\ne{j'}=1}^K 
        \sum_{i'_j=1}^{n_{j}}\sum_{i{''}_{j'}=1}^{n_{j'}} \omega_p(s^{(j)}_{i'_j}) \omega_p(s^{(j')}_{i{''}_{j'}})    \right\}.
\ee
Introducing 
the coefficients $u^{(2)}_{pj}$ defined as 
\be
u^{(2)}_{pj} &= & \int   \omega^2_p(s) p_j(s) ds, 
\ee
the integrals and sums of Eq.~(\ref{eq:Wp2Detail}) reduce to
\be
\la W_p^2\ra&=&   \sum_{j=1}^K \la N_j\ra  \varepsilon_j u^{(2)}_{pj}  + 
          \sum_{j=1}^K \la N_j(N_j-1)\ra \varepsilon^2_j  \left( u_{pj}\right)^2
          + \sum_{j\neq j'=1}^K \la N_jN_{j'}\ra \varepsilon_j\varepsilon_{j'} u_{pj} u_{p{j'}}.
\ee
Note that if terms in equal powers of $N_j$ are regrouped, as in Ref.~\cite{Gorenstein:2011hr}, one ends up with a term in $\la N_j\ra$ with a coefficient proportional to a sum of linear and quadratic powers of the efficiency. It is thus more appropriate to keep the  above expression as is, given it is the factorial moments that are required for the calculation of $\nu_{dyn}$ and they feature a simple square dependence on the detection efficiency. 

The calculation of the covariance $\la W_pW_q\ra$ proceeds in a similar fashion. Introducing  functions, $u_{pqj}$, defined as 
\be
u_{pqj} &= & \int \omega_p(s) \omega_q(s) p_j(s)   ds, 
\ee
one obtains 
\be 
\la W_pW_q\ra & = & 
\sum_{j=1}^K \la N_j\ra \varepsilon_j u_{pqj} 
+ \sum_{j=1}^K \la N_j(N_j-1)\ra u_{pj} u_{qj} \varepsilon_j^2
+ \sum_{j\neq {j'}=1}^K \la N_j N_{j'}\ra \varepsilon_j \varepsilon_{j'}  u_{pj} u_{q{j'}}.
\ee
We thus have obtained formula that express the second moments $\la W_p^2\ra$ and cross-moments $\la W_pW_q\ra$ in terms of the moments $\la N_p\ra$, $\la N_p(N_p-1)\ra$, and $\la N_p N_q)\ra$ in the presence of particle losses with efficiencies $\varepsilon_p$ and $\varepsilon_q$. We now seek to invert these expressions to obtain formula for the moments 
$\la N_p(N_p-1)\ra$  and $\la N_p N_q)\ra$ in terms of $\la W_p^2\ra$ and $\la W_pW_q\ra$.
Proceeding as in Eq.~(\ref{eq:Nmom}), one can absorb the efficiencies into the first moments, second order factorial moments, and covariance by defining
\be \nonumber
\la N_p'\ra & \equiv & \la N_p\ra \varepsilon_p, \\ \label{eq:effCorrMoments}
\la N_p(N_p-1)'\ra & \equiv & \la N_p(N_p-1)\ra \varepsilon_p^2,  \nonumber \\
\la N'_p N'_q\ra & \equiv & \la N_p N_{q}\ra \varepsilon_p^2.
\ee
Expressions for the  moments  $\la W_p^2\ra$ and  $\la W_pW_q\ra$ thus reduce to
\be\label{eq:W2}
\la W_p^2\ra&=&   \sum_{j=1}^K \la N'_j\ra  u^{(2)}_{pj}  + 
          \sum_{i=1}^K \la N_j(N_j-1)'\ra  \left( u_{pj}\right)^2
          + \sum_{j\neq j'=1}^K \la N'_jN'_{j'}\ra u_{pj} u_{p{j'}}, \\
 \label{eq:covWW}
\la W_pW_q\ra & = & 
\sum_{j=1}^K \la N'_j\ra  u_{pqj} 
+ \sum_{j=1}^K \la N_j(N_j-1)'\ra u_{pj} u_{qj} 
+ \sum_{j\neq {j'}=1}^K \la N'_j N'_{j'}\ra u_{pj} u_{q{j'}},
\ee
which defines a system of $(K^2 + K)/2$ linear equations. Proceeding similarly as in Ref.~\cite{Gorenstein:2011hr}, we first define two  ``b'' coefficients
\be\label{eq:bcoeffs}
b_p &=& \la W_p^2\ra - \sum_{j=1}^K u^{(2)}_{pj} \la N'_j\ra, \\ 
b_{pq} &=& \la W_pW_q\ra - \sum_{j=1}^K u_{pqj} \la N'_j\ra, 
\ee
with $p<q$, 
and four sets of ``a'' coefficients 
\be\label{eq:acoeffs}
a^j_p &=& \left( u_{pj}\right)^2, \hspace{0.65in}  1\le p,j \le K; \\  \nonumber 
a^{jj'}_{p} &=& 2 u_{pj}u_{pj'},  \hspace{0.58in}   1\le p \le K; {\rm \ \ \ } 1 \le j<j' \le K; \\ \nonumber 
a^j_{pq} &=&  u_{pj}u_{qj},  \hspace{0.70in}   1\le p < q\le K; {\rm \ \ \ } 1 \le j \le K; \\ \nonumber 
a^{jj'}_{pq} &=&  u_{pj}u_{qj'} + u_{pj'}u_{qj},  \hspace{0.15in}   1\le p < q\le K; 1 \le j < j' \le K.
\ee
We next define the $K+K(K-1)/2$-vectors $\mathbf{N}$ and $\mathbf{B}$  as 
\be
{\mathbf{N}} = \left( {\begin{array}{*{20}{c}}
  {\left\langle {{N_1}({N_1} - 1)'} \right\rangle } \\ 
   \vdots  \\ 
  {\left\langle {{N_K}({N_K} - 1)'} \right\rangle } \\ 
  {\left\langle {{N_1}'{N_2}'} \right\rangle } \\ 
   \vdots  \\ 
  {\left\langle {{N_{k - 1}}'{N_K}'} \right\rangle } 
\end{array}} \right),{\text{     }}{\mathbf{B}} = \left( {\begin{array}{*{20}{c}}
  {{b_1}} \\ 
   \vdots  \\ 
  {{b_K}} \\ 
  {{b_{12}}} \\ 
   \vdots  \\ 
  {{b_{(K - 1)K}}} 
\end{array}} \right)
\ee
and the $(K+K(K-1)/2) \times (K+K(K-1)/2)$ matrix $\mathbf{A}$ as
\be
{\mathbf{A}} = \left( {\begin{array}{*{20}{c}}
  {a_1^1}& \cdots &{a_1^K}&{a_1^{12}}& \cdots &{a_1^{(K - 1)K}} \\ 
   \vdots & \ddots & \vdots & \vdots & \ddots & \vdots  \\ 
  {a_K^1}& \cdots &{a_K^K}&{a_K^{12}}& \cdots &{a_K^{(k - 1)k}} \\ 
  {a_{12}^1}& \cdots &{a_{12}^K}&{a_{12}^{12}}& \cdots &{a_{12}^{(K - 1)K}} \\ 
   \vdots & \ddots & \vdots & \vdots & \ddots & \vdots  \\ 
  {a_{12}^K}& \cdots &{a_{(K - 1)K}^K}&{a_{(K - 1)K}^{12}}& \cdots &{a_{(K - 1)K}^{(K - 1)K}} 
\end{array}} \right).
\ee
Eqs. (\ref{eq:W2},\ref{eq:covWW}) may then be written 
$
{\mathbf{A}}{\mathbf{N}} = {\mathbf{B}}
$,
which is solved by inversion of ${\mathbf{A}}$:
\be\label{eq:NmomSolution}
{\mathbf{N}} = {\mathbf{A}^{-1}} {\mathbf{B}}.
\ee
Note that  while this expression is of the same form as that obtained in 
Ref.~\cite{Gorenstein:2011hr}, the definitions of both ${\mathbf{N}}$ and ${\mathbf{B}}$ are quite different and the procedure outlined in this work is thus distinct from the original identity method.  

Three remarks are in order. First, since the calculation of $b_p$ and  $b_{pq}$ requires 
knowledge of $\la N'_j\ra$, one must first solve Eq.~(\ref{eq:WpFirstMom}) before attempting the solution of Eq.~(\ref{eq:NmomSolution}). Second, once the moments $\la N_p(N_p - 1)' \ra$ and $\la N'_p N'_q\ra$ are obtained from Eq.~(\ref{eq:NmomSolution}), it is then 
unnecessary to correct them for efficiencies towards the determination of  $\nu_{dyn}$
using 
\be\label{eq:nudyn2}
\nu_{dyn} = \frac{\la N_p\left( N_p-1\right)'\ra }{\la N'_p\ra^2 } +  \frac{\la N_q\left( N_q-1\right)'\ra }{\la N'_q\ra^2 } - 2  \frac{\la N'_p N'_q\ra }{\la N'_p\ra \la N'_q\ra },
\ee
since the efficiencies cancel out term by term in this expression. Finally, it should be clear that for the purpose of a measurement of $\nu_{dyn}$, the identity method as formulated in Ref.~\cite{Gorenstein:2011hr} shall produce proper results because the method is linear and thus produces ratios $\la n(n-1)\ra/\la n\ra^2$ that are robust even though the  moments  $\la n^2\ra$ feature a non factorizable dependence on the detection efficiency.  
However, the method outlined in this section presents the advantage of yielding factorial moments $\la n(n-1)\ra$ which have a simpler dependence on the efficiency, and it is straighforward, as we show in the following section,  to extend the method to account for efficiency dependencies on the particle momentum or direction.

\section{The Identity Method with several $p_{\rm T}$ bins}
\label{sec:IMwithPtBins}

The method outlined in the previous section assumes that the line shape of the PID signal $s$  is independent of the momentum and direction of the particles. In practice, for instance, the energy loss of a particle does depend on its momentum and the $dE/dx$ line shape is then  a function of the particle  momentum. This in turn implies that the probabilities $p_p(s)$ are  also dependent on the momentum of the particles. The identity method analysis must then be carried out in fine bins of momentum and one must  also consider  how detection efficiencies may change with the particle momentum and direction (i.e., vs. $p_{\rm T}$, rapidity, and azimuth angle). The calculation technique used in the previous section remains applicable provided one assumes there is a definite (albeit unknown a priori) probability to find particles in specific bins of  $p_{\rm T}$, rapidity, and azimuth angle. In the following, we carry out the calculation with finite momentum binning exclusively, but the technique can be extended to account for binning in other coordinates. 

In order to account for particle production in $R$ momentum bins, we replace the probability distribution  $P_{\rm M}(n_1,n_2,\ldots,n_K)$ by a new function  $P_{\rm M}(n_{11},n_{12},\ldots,n_{1R} ,n_{21},\ldots,n_{2m_p},\ldots,n_{K1},\ldots,n_{KR})$, in which the variables $n_{i\alpha}$, with $i=1,\ldots, K$, $\alpha=1,\ldots, R$,  denote the number of particles of species $i$ produced in momentum bin $\alpha$. Hereafter, we use roman letters to index particle species and greek letters to index momentum bins. Equations \ref{eq:MeasMultMoments}  must  
be extended to include momentum bin dependencies. Introducing the shorthand 
\be
\vec n \equiv (n_{11},n_{12},\ldots,n_{1R} ,n_{21},\ldots,n_{2R},\ldots,n_{K1},\ldots,n_{KR}),
\ee
as well as the sum notation
\be
\sum_{\vec n} \equiv \sum_{n_{11}=0}^{\infty}\cdots \sum_{n_{1R}=0}^{\infty}
\sum_{n_{21}=0}^{\infty}\cdots \sum_{n_{2R}=0}^{\infty} \cdots 
\sum_{n_{K1}=0}^{\infty}\cdots \sum_{n_{KR}=0}^{\infty},
\ee
the moments of the multiplicities $n_{i\alpha}$  can  be calculated for each species $p$ and each $p_{\rm T}$ bin $\alpha$, according to
\be\nonumber
\la n_{p\alpha}\ra &\equiv & \sum_{\vec n} P_{\rm M}(\vec n) n_{p\alpha}, \\ \label{eq:ExtendedMeasMultMoments}
\la n_{p\alpha}^2\ra &\equiv &  \sum_{\vec n} P_{\rm M}(\vec n) n_{p\alpha}^2, \\ \nonumber
\la n_{p\alpha} n_{q\beta}\ra &\equiv &  \sum_{\vec n} P_{\rm M}(\vec n) n_{p\alpha} n_{q\beta}. 
\ee
For fluctuations analyses, one seeks the moments of multiplicities $n_{p}$ consisting of the sum of the    $n_{p\alpha}$ across all $p_{\rm T}$ bins, i.e.,
\be
n_{p} = \sum_{\alpha=1}^{R} n_{p\alpha}.
\ee
The first moment $\la n_{p}\ra$  is trivially obtained as a sum of the  first moments $\la n_{p\alpha}\ra$ 
\be
\la n_{p}\ra = \lla \sum_{\alpha=1}^{R} n_{p\alpha}\rra = \sum_{\alpha=1}^{R} \la n_{p\alpha}\ra
\ee
Second moments and covariances require one sums all relevant momentum bin combinations
\be  
\la n_{p}^2\ra &=&\lla \left( \sum_{\alpha=1}^{R}  n_{p\alpha} \right)^2 \rra  =  \sum_{\alpha=1}^{R}  \lla n_{p\alpha}^2 \rra + \sum_{\alpha\ne\alpha'=1}^{R}  \lla n_{p\alpha} n_{p\alpha'} \rra  \\ 
\la n_{p}n_{q}\ra &=&\lla \left( \sum_{\alpha=1}^{R}  n_{p\alpha} \right)  \left( \sum_{\alpha'=1}^{R}  n_{q\alpha'} \right)\rra  =  \sum_{\alpha,\alpha'=1}^{R}  \lla n_{p\alpha} n_{q\alpha'} \rra.
\ee
Evidently, our discussion of efficiency losses applies for each momentum bin individually. One can then write
\be 
\la n_{p\alpha} \ra &=& \varepsilon_{p\alpha} \la N_{p\alpha} \ra \\  \label{eq:ExtendedMeasVsProdMoments}
\la n_{p\alpha}\left( n_{p\alpha} - 1\right)\ra &=&  \varepsilon_{p\alpha}^2 \la N_{p\alpha}\left( N_{p\alpha} - 1\right)\ra \\
\la n_{p\alpha} n_{q\beta}\ra &=& \varepsilon_{p\alpha} \varepsilon_{q\beta} \la N_{p\alpha} N_{q\beta} \ra,
\ee
where the variables $n_{p\alpha}$ and $N_{p\alpha}$ represent the measured and true numbers of particles of species $p$ in momentum bin $\alpha$, respectively.
A proper calculation of the moments $\la N_{p}\ra$,  $\la N_{p}\left(N_{p}-1 \right)\ra$ and $\la N_{p} N_{q} \ra$ shall then require efficiency corrections $p_{\rm T}$-bin by $p_{\rm T}$-bin, if the efficiencies 
$\varepsilon_{p\alpha}$ depend on $\alpha$, i.e., the momentum of the particles.
\be\label{eq:NpAvg}
\la N_{p}\ra &=&  \sum_{\alpha=1}^{R} \frac{ \la n_{p\alpha}\ra}{\varepsilon_{p\alpha}} \\  \label{eq:NpNpAvg}
\la N_{p}\left(N_{p}-1\right) \ra &=& \sum_{\alpha=1}^{R}  \frac{\lla n_{p\alpha}\left(n_{p\alpha}-1\right) \rra}{\varepsilon_{p\alpha}^2}  + \sum_{\alpha\ne\alpha'=1}^{R}  \frac{\lla n_{p\alpha} n_{p\alpha'} \rra}{\varepsilon_{p\alpha}\varepsilon_{p\alpha'}}  \\ \label{eq:NpNqAvg}
\la N_{p}N_{q}\ra &=&  \sum_{\alpha,\alpha'=1}^{R}  \frac{\lla n_{p\alpha} n_{q\alpha'} \rra}{\varepsilon_{p\alpha}\varepsilon_{q\alpha'}}.
\ee
Equations (\ref{eq:NpAvg}-\ref{eq:NpNqAvg}) are general and can be applied to  traditional cut analyses or with the $p_{\rm T}$  identity method we discuss next. 

To apply the identity method in cases involving multiple $p_{\rm T}$ bins, one must 
obtain expressions for the moments $\la n_{p\alpha}\ra$, $\lla n_{p\alpha}\left(n_{p\alpha}-1\right) \rra$, and $\lla n_{p\alpha} n_{q\alpha'} \rra$ in terms of identity variables determined for each species and each momentum bin. We thus define 
\be
W_{p\alpha} = \sum_{i=1}^n \omega_{p\alpha}(s_i) \Theta_{\alpha}(s_i),
\ee
in which the sum proceeds over all (accepted) particles of an event. The function   $\omega_{p\alpha}(s_i)$ represents the probability of the $i$-th particle being of species $p$ when observed in $p_{\rm T}$ bin $\alpha$, and the function $\Theta_{\alpha}(s_i)$ is unity if the $i$-th particle is within the $p_{\rm T}$ bin $\alpha$ and null otherwise. Calculations of the expectation value of the moments of $W_{p\alpha}$ proceed  as in sec.~\ref{sec:IdentityMethod} but are carried out for specific   $p_{\rm T}$ bins $\alpha$ ($\beta$). The first moments are 
\be
\la W_{p\alpha}\ra = \sum_{j=1}^{K} \la n_{p\alpha}\ra u_{pj,\alpha} = \sum_{j=1}^{K} \la N_{p\alpha}\ra u_{pj,\alpha} \varepsilon_{i\alpha},
\ee
in which the coefficients $u_{pj,\alpha}$ are calculated according to 
\be
u_{pj,\alpha} = \int  \omega_{p\alpha}(s) p_{j\alpha}(s) ds,
\ee
where $p_{j\alpha}(s)$ represents the probability of observing a PID signal $s$  for a particle of species $j$ in momentum bin $\alpha$. 

Four second order moments must be considered which we denote $\la W_{p\alpha}^2\ra$, $\la W_{p\alpha}W_{p\beta}\ra$,   $\la W_{p\alpha}W_{q\alpha}\ra$,  and $\la W_{p\alpha}W_{q\beta}\ra$,
with  $p < q$ and  $\alpha \ne \beta$. Calculation of these moments yields 
\be\label{eq:WmomentsWithPtBins1}
\la W_{p\alpha}^2\ra &=& \sum_{j=1}^K \la N_{j\alpha} \ra \varepsilon_{j\alpha} u^{(2)}_{pj,\alpha}
+ \sum_{j=1}^K \la N_{j\alpha} (N_{j\alpha}-1)\ra \varepsilon_{j\alpha}^2 \left(u_{pj,\alpha}\right)^2
+ \sum_{j\ne j'=1}^K \la N_{j\alpha} N_{j'\alpha}\ra \varepsilon_{j\alpha}\varepsilon_{j'\beta}    u_{pj,\alpha}u_{pj',\alpha} \\ \label{eq:WmomentsWithPtBins2}
\la W_{p\alpha}W_{p\beta}\ra &=& 
\sum_{j,j'=1}^K \la N_{j\alpha} N_{j'\beta}\ra \varepsilon_{j\alpha}\varepsilon_{j'\beta}    u_{pj,\alpha}u_{pj',\beta} \\ \label{eq:WmomentsWithPtBins3}
\la W_{p\alpha}W_{q\alpha}\ra &=& 
\sum_{j=1}^K \la N_{j\alpha}\ra \varepsilon_{j\alpha}  u_{pqj,\alpha} 
+\sum_{j=1}^K \la N_{j\alpha}(N_{j\alpha}-1)\ra \varepsilon_{j\alpha}^2  u_{pj,\alpha}u_{qj,\alpha}
+\sum_{j\ne j'=1}^K \la N_{j\alpha}N_{j'\alpha}\ra \varepsilon_{j\alpha}\varepsilon_{j'\alpha} u_{pj,\alpha}u_{qj',\alpha} \\  \label{eq:WmomentsWithPtBins4}
\la W_{p\alpha}W_{q\beta}\ra &=& \sum_{j,j'=1}^K \la N_{j\alpha} N_{j'\beta}\ra \varepsilon_{j\alpha}\varepsilon_{j'\beta}    u_{pj,\alpha}u_{qj',\beta}.
\ee
where we introduced the  coefficients
\be
u^{(2)}_{pj,\alpha} &= & \int \omega_{p\alpha}(s)^2 p_{j\alpha}(s)   ds,  \\ 
u_{pqj,\alpha} &= & \int \omega_{p\alpha}(s) \omega_{q\alpha}(s) p_{j\alpha}(s)   ds.
\ee
By construction, the cross-terms are symmetric under interchanges of the 
indices $p$ and $q$ and indices $\alpha$ and $\beta$: 
\be\nonumber 
\la W_{p\alpha}W_{p\beta}\ra &=& \la W_{p\beta}W_{p\alpha}\ra \\   \nonumber
\la W_{p\alpha}W_{q\alpha}\ra &=& \la W_{q\alpha}W_{p\alpha}\ra \\  \nonumber
\la W_{p\alpha}W_{q\beta}\ra &=& \la W_{q\beta}W_{p\alpha}\ra.
\ee
 There are thus $K\times R$ independent terms of the form $\la W_{p\alpha}^2\ra$,  
$K\times R(R-1)/2$   of the form $\la W_{p\alpha}W_{p\beta}\ra$,
$K(K-1)/2 \times R$  of the form $\la W_{p\alpha}W_{q\alpha}\ra$,
and $K(K-1)/2 \times R(R-1)/2$  of the form $\la W_{p\alpha}W_{q\beta}\ra$.
The relation between the second order moments of $W_p$ and the second order moments of 
the multiplicities $N_p$ may then be viewed as a system of  
  $Q= (K+ K(K-1)/2) \times (R+R(R-1)/2)$ independent linear equations.

Proceeding as in sec.~\ref{sec:IdentityMethod}, we define ``b'' coefficients according to 
\be\label{eq:bcoeffsWithPt}
b_{p,\alpha\alpha} &=& \la W_{p\alpha}^2\ra - \sum_{j=1}^K \la N_{p\alpha}\ra \varepsilon_{j\alpha} u^2_{pj,\alpha} , \\ 
b_{p,\alpha\beta} &=& \la W_{p\alpha}W_{p\beta}\ra, \\ 
b_{pq,\alpha\alpha} &=& \la W_{p\alpha}W_{q\alpha}\ra - \sum_{j=1}^K \la N_{j\alpha}\ra \varepsilon_{j\alpha} u_{pqj,\alpha}, \\
b_{pq,\alpha\beta} &=& \la W_{p\alpha}W_{q\beta}\ra,
\ee
where $p<q$ and $\alpha \ne\beta$.
The ``a'' coefficients are next defined according to 
\be
a_{p,\alpha}^{j} &=& \left( u_{pj,\alpha}\right)^2 \varepsilon^2_{j\alpha} \\ 
a_{p,\alpha}^{jj'} &=& u_{pj,\alpha} u_{pj',\alpha} \varepsilon_{j\alpha} \varepsilon_{j'\alpha} \\ 
a_{pq,\alpha}^j &=&  u_{pj,\alpha} u_{qj,\alpha} \varepsilon^2_{j\alpha}  \\ 
a_{pq,\alpha}^{jj'} &=&  u_{pj,\alpha} u_{qj',\alpha} \varepsilon_{j\alpha}\varepsilon_{j'\alpha} \\
a_{pq,\alpha\beta}^{jj'} &=&  u_{pj,\alpha} u_{qj',\beta} \varepsilon_{j\alpha}\varepsilon_{j'\beta}. 
\ee

The column vector $\vec B$, matrix $\vec A$, and column vector $\vec N$ may then written
\be
{\mathbf{B}} = \left( {\begin{array}{*{20}{c}}
  {{b_{p,\alpha\alpha }}} \\ \\ 
  {{b_{p,\alpha\beta }}} \\  \\ 
  {{b_{pq,\alpha \alpha }}} \\ \\ 
  {{b_{pq,\alpha \beta  }}} 
 \end{array}} \right) 
{\text{    }}
{\mathbf{A}} = \left( {\begin{array}{*{20}{c}}
  a_{p,\alpha}^j & a_{p,\alpha}^{jj'} &  0 \\  \\
  0 & 0 &  a_{p,\alpha\beta}^{jj'} \\  \\
  a_{pq,\alpha\alpha}^{j} & a_{pq,\alpha\alpha}^{jj'} &  0 \\  \\
  0 & 0 &   a_{pq,\alpha\beta}^{jj'}   
  \end{array}} \right) 
{\text{    }}
{\mathbf{N}} = \left( {\begin{array}{*{20}{c}}
  {\la N_{j,\alpha}(N_{j,\alpha}-1)\ra } \\ \\ 
  {\la N_{j,\alpha}N_{j',\alpha}\ra } \\ \\
  {\la N_{j,\alpha}N_{j',\beta}\ra } 
\end{array}} \right) 
\ee
in which each of the elements are themselves vectors or matrices with indices $p$, $q$ spanning all values $1\le p< q  \le K$ and  indices $\alpha$ and $\beta$ spanning all values  $1\le \alpha\ne \beta \le R$. For instance, in the case of $b_{p,\alpha}$, $\alpha$ spans all values $1$ to $R$ while 
$p$  spans all values from 1 to $K$. However, in the case of the other $b$ coefficients, the values 
spanned should satisfy $\alpha\ne \beta$ and $p<q$.  
Equations~(\ref{eq:WmomentsWithPtBins1}-\ref{eq:WmomentsWithPtBins4}) may then be written $\mathbf{B} = \mathbf{A} \mathbf{N}$
and can be solved by inversion of the matrix $\mathbf{A}$:
\be
\mathbf{N} = \mathbf{A}^{-1} \mathbf{B}
\ee
It is important to note  that both $\mathbf{A}$ and $\mathbf{B}$ are now explicitly dependent on the detection efficiencies $\varepsilon_{j\alpha}$. Given the efficiencies are $p_{\rm T}$ dependent, efficiency coefficients must be indeed  included explicitly in the expressions of $\mathbf{A}$ and $\mathbf{B}$. The robustness of ratios $\la N_{j,\alpha}(N_{j,\alpha}-1)\ra/\la N_{j,\alpha}\ra^2$ is thus effectively lost. The identity method remains nonetheless applicable provided the coefficients $u_{pj,\alpha}$, $u^{(2)}_{pj,\alpha}$, $u_{pqj,\alpha}$, and the efficiencies $\varepsilon_{j\alpha}$ can be evaluated with sufficient precision.

\section{Summary}
\label{sec:Summary}

We first discussed the impact of finite particle losses associated with instrumental effects in  measurements of moments of produced multiplicities with the Identity Method towards the evaluation of fluctuation measures such as $\nu_{dyn}$. We found that  the original  identity
method produces moments $\la n^2\ra$ with a complex dependence on the detection 
efficiency while the procedure outlined in this work yields factorial moments $\la n(n-1)\ra$ that feature a simple square dependence on the efficiency. However, both the 
original and modified identity methods shall yield robust, i.e., efficiency independent results, for the fluctuation observable $\nu_{dyn}$ as long as particle detection efficiencies are momentum independent.
We further showed that the modified method outline in this work provides for a straightforward albeit somewhat tedious extension to  experimental cases where detection efficiencies are strongly dependent on the momentum of particles.  

The treatment of particle losses discussed in this work can and should be applied to measurements of higher moments discussed in Ref.~\cite{Rustamov:2012bx}. 

\newenvironment{acknowledgement}{\relax}{\relax}
\begin{acknowledgement}
\section*{Acknowledgements}

The author thanks colleagues S.~Voloshin and A.~Rustamov for fruitful discussions and comments.

\end{acknowledgement}

\bibliography{imr}

\end{document}